\documentclass[aps,nofootinbib,endfloats]{revtex4}
\usepackage{epsfig}
\input epsf.tex
\usepackage[english]{babel}
\usepackage{color}

\usepackage{graphicx}
\usepackage{amsmath}

\begin{document}
\draft
\title{Optical Properties of Rydberg Excitons and Polaritons}
\author{Sylwia Zieli\'{n}ska-Raczy\'{n}ska,
Gerard Czajkowski, and David Ziemkiewicz} \affiliation{ Institute
of Mathematics and Physics, UTP University of Science and
Technology, Al. Prof. S. Kaliskiego 7, PL 85-789 Bydgoszcz
 (Poland)}

\begin{abstract} We show how to compute the optical functions when Rydberg Excitons
appear, including the effect of the coherence between the
electron-hole pair and the electromagnetic  field. We use the Real
Density Matrix Approach (RDMA), which, combined with Green's
function method, enables one to derive analytical expressions for
the optical functions. Choosing the susceptibility, we performed
numerical calculations appropriate to a  Cu$_2$0 crystal. The
effect of the coherence is displayed in the line shape. We also
examine in details and explain the dependence of the oscillator
strength and the resonance placement on the state number. We
report a good agreement with recently published experimental
data.\end{abstract} \maketitle
\section{Introduction}
Since the 1980s Rydberg atoms, in which the valence electron is in
a state of high principal quantum number $n\gg 1$, have been
extensively studied. They have exaggerated atomic properties
including dipole-dipole interactions that scale as $\sim n^4$ and
radiative lifetimes proportional to $n^2$. Another important
property of the Rydberg states is the large orbital radius, and
hence dipole moment $\sim n^2$.  The natural consequence of the
large dipole moment featured by the Rydberg atoms is the large
interaction between two of them - one is able to observe
dipole-dipole interactions between atoms on the $\mu$ scale.
Another consequence of the incredibly large dipole moment is an
exaggerated response to external fields \cite{Gallagher}. Due to a
small admixing of the Rydberg state with the ground one, the atoms
experience long-range and large interactions while keeping a long
lifetime in a quantum superposition, which is  very useful for
creating qubits, so
 Rydberg atoms are well suited to  applications in quantum information processing.
The idea of using dipolar Rydberg interactions to implement the
 Rydberg blockade bases on the fact that in an ensemble of atoms with  long-range dipolar interactions between them, only one atom can be excited at given time. This 'dipole blockade' has now been observed for two single atoms positioned at macroscopic distances \cite{Weide}.
 Rydberg interaction between distanced atoms is used to implement one or two-qubit quantum gates of high fidelity \cite{qinf}.
 Coupling of Rydberg states results in d.c. Kerr
 effect which is six order of magnitude greater then in conventional Kerr.
 This result has great impact on development of high-precision electric field sensors and other nonlinear optical devices \cite{nat}.
 Rydberg atoms are also applied in single-microwave-photon counters what
 is especially promising in high precession measurements in quantum optics \cite{ver}.
 On this wide background of potential applications established in quantum optics and atomic physics demonstration of
 gigant Rydberg excitons in cooper oxide (\cite{Hofling}, \cite{Kazimierczuk}) is a great step
 toward studies that are out of the reach in atomic physics, f.e. Rydberg excitons, due
 to their smaller energies, need lower magnetic field to mimick hydrogen atoms
 in white-draft stars \cite{fried} or they are promising candidate for creating Bose-Einstein condensat in solid~\cite{Stolz}.

The phenomenon of excitons and its consequences on the optical
spectra of semiconductors, intensively studied over the decades,
obtained a new impulse when  the so-called Rydberg excitons have
been detected in a natural crystal of copper oxide found at the
Tsumeb mine in Namibia \cite{Hofling}, \cite{Kazimierczuk}. In the
simplest picture the exciton (we have in mind the so-called
Wannier exciton) is modelled as a hydrogen-like atom, composed
from the electron and the hole, interacting via the Coulomb
potential screened by the semiconductor dielectric constant. The
materials, where Wannier excitons occur, show optical spectra
where the transition states, related to the principal quantum
number $n$, are observed.  In most previously studied materials,
like, for example, GaAs, only few excited excitonic states
($n=1,2,3$) were detected, which was caused by the small excitonic
binding energy (in the order of a few meV, as in GaAs), and
dissipative processes. Therefore the discovery reported in Ref.
\cite{Kazimierczuk}, where the analysis of the spectra revealed
spectral absorption lines associated with the formation of
excitons that have principal quantum numbers as large as n = 25,
provoke a new situation in the condensed matter optics. Many
problems as, for example, the light-matter interaction when the
excitons with such large quantum numbers are present, will expect
new description. Excitonic spectra of highly excited Rydberg
excitons observed recently resemble that of a hydrogen atom
consisting of a Rydberg series but even
 if the hydrogen picture of the exciton seems
to be very simple, then, the calculations with eigenfunctions
related to $n=25$ would be not trivial. The standard description
needs to be revised do to the fact that a size of the exciton is
much larger then the wavelenght of light used to create it
\cite{Hofling}.
 There are many factors,
as, for example, the band structure, the temperature, the laser
power, the dissipative processes, which should be included in the
theoretical description.

Here we neither  enter into the analysis of the experiment nor in
the calculations of the band structure. We propose a method, which
gives a simple expression for the optical functions, taking into
account excitonic states of arbitrary order which allows one to
obtain theoretical spectra and to analyze the experimental ones.
The method is based on the Real Density Matrix Approach and uses
Green's function method to solve the Schr\"{o}dinger-like
equations which are typical for this approach. The method will
also give insight into the aspect of polaritons, which are closely
related to the excitonic states. The RDMA, initiated by the works
by Stahl \emph{et al}.
 see, for example, \cite{StB87}, \cite{CBT96}, was very
 succsessful in describing optical properties of semiconductors
 for energies near the fundamental gap, where the excitons are
 relevant. This approach also solved the old ABC problem (for example, \cite{Birman82}-\cite{Agran2009}), at least
 for the cases with a few excitonic states \cite{CBT96}. In what follows we focuss the attention on the
 optical spectra of Cu$_2$O. As it follows by the analysis of crystal symmetry, the lines related
 to odd angular momentum exciton number $\ell=1,3,..$ are observed \cite{Thewes}. The dominant role play the $P$-excitons
 (the so-called yellow series), but also
 excitonic states with higher than $\ell=1$ angular momentum  (for example, the $F$- excitons with $\ell=3$ and $H$- excitons with $\ell=5$) were
 observed in one-photon
absorption spectra of high-quality  cuprous oxide.  Our method
gives not only the energy eigenvalues, but also the line shapes of
the optical functions, from which we have chosen the
susceptibility. The presented theory explains many peculiar
characteristics of Rydberg excitons such as deviations from
$n^{-3}$ law of oscillator strengths and $n^{-2}$ law for the
excitonic energies and gives the polariton dispersion relation.
Using anisotropic effective masses, we show the energy splitting
of the $P$, $F$, and $H$ excitons.  Our numerical results are in
agreement with measurements obtained recently in the outstanding
experiments performed by Kazimierczuk \emph{et al.
}\cite{Kazimierczuk} and Thewes \emph{et al.} \cite{Thewes}.

 It is believed that the observation of Rydberg excitons allows one to open a new field in condensed matter spectroscopy.
 For highly excited Rydberg excitons in Cu$_2$O the scale is of over 1$\mu$m \cite{Tizei} so
 the application of solid state huge-size excitons as all-optical switching, mesoscopic single-photon devices or their implementation to
 construction of quantum gates influences the development of new experimental technics.
 Additionally, it is interesting to note that the excitonic approach toward the old idea of Rydberg atoms is example how the development
 of one field inspires the others.

 Our paper is organized as follows. In the section \ref{density.matrix},
 we briefly recall the basic equations of the RDMA approach  and derive expressions for the interband susceptibility.
  Next, in section
\ref{Cu2Osusceptibility}, the derived expression is analyzed for
the case of Cu$_2$O crystal.
 In section \ref{polariton.dispersion},
  the derived expression for the susceptibility is used to obtain the polariton dispersion relation of the considered cuprous oxide
  crystal. In section \ref{anisotropy} we show the impact of the
  effective masses anisotropy on the calculated optical
  properties.
In section \ref{Green} the method of calculating the
susceptibility and the polariton dispersion in terms of an
appropriate Green's function is presented.
 In section \ref{results} the results for the absorption spectra are presented and discussed.
   The comparison of obtained results with experimental data and a brief overview of optimizing procedure is
   included.We close with final remarks in section \ref{finalremarks}.

\section{Density matrix formulation}\label{density.matrix}
Having in mind the experiments by Kazimierczuk \emph{et
al}.~\cite{Kazimierczuk}, we will compute the linear optical
response of a semiconductor platelet to a plain electromagnetic
wave
\begin{equation}\label{stpar_wave}
\textbf{E}_i(z,t)=\textbf{E}_{i0}\exp({\rm i}k_0z-{\rm i}\omega
t), \qquad k_0=\frac{\omega}{c},
\end{equation}
\noindent attaining the boundary surface located at the plane
$z=0$. The second boundary is located at the plane
 $z=L$. As indicated above, we use the RDMA. In the linear case,
 the optical response is obtained by solving the so-called
 constitutive equations, supplemented by the Maxwell equations for
 the wave propagating in the semiconductor crystal. Considering a
 semiconductor with a nondegenerate conduction band and a
 $\lambda-\hbox{fold}$ degenerate valence band the constitutive equations
 have the form (for example, \cite{StB87})
 \begin{equation}\label{constitutiveeqn}
 \dot{Y}(\textbf{R},\textbf{r})=(-{\rm
 i}/\hbar)H^\lambda_{eh}{Y}(\textbf{R},\textbf{r})-{\mit\Gamma}^\lambda{Y}(\textbf{R},\textbf{r})+({\rm
 i}/\hbar)\textbf{E}(\textbf{R})\textbf{M}^\lambda(\textbf{r}),
 \end{equation}
where $Y^\lambda$ is the bilocal coherent electron-hole amplitude
(pair wave function), ${\bf R}$ jest is the excitonic
center-of-mass coordinate, $\textbf{r}=\textbf{r}_e-\textbf{r}_h$
the relative coordinate, $\textbf{M}^\lambda(\textbf{r})$ the
smeared-out transition dipole density, and ${\bf E}({\bf R})$ is
the electric field vector of the wave propagating in the crystal.
The smeared-out transition dipole density ${\bf M}({\bf r})$ is
related to the bilocality of the amplitude $Y$ and describes the
quantum coherence between the macroscopic electromagnetic field
and the interband trasitions. Its form depends on the type of the
interband transition (direct or indirect gap) and will be
specified below. The two-band Hamiltonian $H_{eh}$ is taken in the
form
\begin{eqnarray}\label{hamiltonians}
\begin{array}{llll}
H^\lambda_{eh}&=& H^\lambda_{c.m}+H^\lambda_r,&\\
H^\lambda_{c.m}&=&(-\hbar^2/2)\nabla_R({\underline{\underline
M}^\lambda})^{-1}\nabla_R+\hbar\omega_g^\lambda,&\hbox{(center-of-mass motion)}\\
H_r&=&(-\hbar^2/2)\nabla_r({\underline{\underline
\mu}}^\lambda)^{-1}\nabla_r+V_{eh}(r),&\hbox{(relative motion)},\\
\end{array}
\end{eqnarray}
${\underline{\underline \mu}^\lambda},{\underline{\underline
M}^\lambda}$ being the exciton reduced- and total mass tensors,
respectively, and $\hbar\omega_g^\lambda$ is the energy gap for
the considered pair of energy levels. Operators
${\mit\Gamma}^\lambda$ stand for irreversible processes. The
coherent amplitudes $Y^\lambda$ define the excitonic counterpart
of the polarization
\begin{equation}\label{polarization}
\textbf{P}(\textbf{R})=2\sum\limits_\lambda \int {\rm d}^3 r~
\hbox{Re}~\left[\textbf{M}^\lambda(r)Y^\lambda(\textbf{R},\textbf{r})\right],
\end{equation}
which is than used in the Maxwell field equation
\begin{equation}\label{Maxwell}
c^2\nabla_R^2
\textbf{E}-\underline{\underline{\epsilon}}_b\ddot{\textbf{E}}(\textbf{R})=\frac{1}{\epsilon_0}\ddot{\textbf{P}}(\textbf{R}),
\end{equation}
with the use  of the bulk dielectric tensor
$\underline{\underline{\epsilon}}_b$ and the vacuum dielectric
constant $\epsilon_0$. Having the polarization, we can compute the
excitonic susceptibility $\underline{\underline{\chi}}(\omega,k)$
\begin{equation}\label{susceptibility}
\textbf{P}(\omega,\textbf{k})=\epsilon_0\underline{\underline{\chi}}(\omega,k)\textbf{E}(\omega,\textbf{k}).
\end{equation}
In the present paper we solve the equations
(\ref{constitutiveeqn})-(\ref{susceptibility}) with the aim to
compute the optical functions (reflectivity, transmission, and
absorption) for the case of Cu$_2$O. The first step is to
calculate the dielectric susceptibility. This can be achieved in
two ways: 1) by expanding the coherent amplitudes $Y^\lambda$ in
terms of eigenfunctions of the Hamiltonian $H_r$ of the relative
electron-hole motion, 2) using the appropriate Green function of
the l.h.s. operator in eq. (\ref{constitutiveeqn}). We begin with
the method 1, for the case of an unbounded semiconductor crystal.
Assume that there exists an orthonormal basis
$\{\varphi_n^\lambda\}$ of eigenfunctions of the operator
$H_r^\lambda$ and $E_n^\lambda$ are the  corresponding
eigenvalues. The eigenfunctions of the total Hamiltonian
$H^\lambda_{eh}$ have the form
\begin{equation}
\Phi^\lambda_{\textbf{k},n}=\exp({\rm i}\textbf{kR})
\varphi_n^\lambda,
\end{equation}
with the corresponding eigenvalues
\begin{equation}\label{omega_n}
\hbar\Omega_n^\lambda(\textbf{k})=\sum\limits_{\alpha=1}^3\frac{\hbar
k_\alpha^2}{2m^\lambda_\alpha}+\hbar\omega_g^\lambda+E_n^\lambda,
\end{equation}
with the assumption that the total effective-mass tensor has a
diagonal form. We also assume that the
$\Phi^\lambda_{\textbf{k},n}$ are eigenfunctions of the damping
operators ${\mit\Gamma}^\lambda$ corresponding to the eigenvalues
$\gamma^\lambda_n(\textbf{k})$. Expanding both $Y^\lambda$ and
$\textbf{M}^\lambda(\textbf{r})\textbf{E}(\textbf{R})$ in terms of
$\Phi^\lambda_{\textbf{k},n}$  and going over to the Fourier
representation, we obtain (\cite{CzChmara})
\begin{equation}\label{susceptibilitychi}
\chi(\omega,\textbf{k})=\frac{1}{\epsilon_0\hbar}\sum\limits_\lambda\sum\limits_n
\left[\frac{c^{*\lambda}_{n\alpha}c^{\lambda}_{n\beta}}{\Omega_n^\lambda(\textbf{k})-(\omega+{\rm
i}\gamma_n^\lambda(\textbf{k}))}+\frac{c^{\lambda}_{n\alpha}c^{*\lambda}_{n\beta}}{\Omega_n^\lambda(\textbf{k})+(\omega+{\rm
i}\gamma_n^\lambda(\textbf{k}))}\right],
\end{equation}
where
\begin{equation}
\textbf{c}^\lambda_n=\langle \varphi^\lambda_n\vert
\textbf{M}\rangle=\int {\rm d}^3 r \;\varphi^{*\lambda}_n(r)
\textbf{M}^\lambda(r).
\end{equation}

\begin{table}[h]
\centering
\caption{\small Two sets of band parameter values for Cu$_2$O from
Refs.~\cite{fried} and \cite{Thewes}, energies in meV, masses in free electron mass
$m_0$, lengths in nm, $\gamma_1, \gamma_2, \gamma_3$ are Luttinger
parameters (here only $\gamma_1$, from Ref.~\cite{Thewes}),
$\alpha$ is the anisotropy parameter, energies $\vert E_{n\ell
m}\vert=\eta_{\ell m}^2R^*/n^2$, $E_{Tn\ell m}$ are positions of
resonances} 

\begin{tabular}{c c c }
\hline\\
Parameter~ & \hbox{[\cite{fried}]}& \hbox{[\cite{Thewes}]} \\
\hline $E_g$ & 2172.08&2172.08\\
 $m_e$ & 0.99 &1.01\\
$\gamma_1$&&$1.79$\\
$m_h[110]=m_{h\parallel}$ &0.66 &0.5587\,$^a$  \\
${m_h[001]=m_{hz}\,}^b$  &2.01&1.99  \\
$\mu[110]=\mu_{\parallel}$ & 0.396 & 0.3597$^c$ \\
$\mu[001]=\mu_{z}$ &0.663& 0.672 \\
$M[110]=M_\parallel$&1.65&1.5687\\
$M[001]=M_z$&3.0&3.0\\
$\alpha=\mu_\parallel/\mu_z$&0.597&0.535\\
$\eta_{00}$&1.0669&1.1004\\
$\eta_{10}$&1.496&1.1901\\
$\eta_{30}$&1.408&1.168\\
$\eta_{50}\;^d$&&1.1172\\
$R^*$&95.74&86.981\\
$\vert E_{100}\vert$&108.95&105.32\\
$\vert E_{210}\vert$&53.56&30.80\\
$\vert E_{410}\vert$&13.39&7.70\\
$\vert E_{430}\vert$&11.86&7.4162\\
$\vert E_{710}$&&2.5247\\
$\vert E_{750}\vert$&&2.2162\\
$\vert E_{730}$ &&2.4222\\
$E_{TP410}$&2158.69&2164.38\\
 $E_{TF430}$&2160.22&2164.67\\
 $E_{TP710}$&&2169.56\\
 $E_{TF730}$&&2169.66\\
 $E_{TH750}$&&2169.86\\
$\Delta E(FP)$&1.53&0.29\\
$a^*$&1.00&1.1\\
$\epsilon_b$&7.5&7.5\\
$\epsilon_\infty$&6.5&6.5\\
\hline\\
$^a$ from $m_h=1/\gamma_1$, $^b\,$ by $M-m_e$ for the total mass $M=3m_0$~\cite{Dasbach}\\
$^c$ from $\mu=1/\gamma_1',~\gamma_1'=\gamma_1+m_0/m_e$\\
$^d$ from $\eta_{\ell m}\approx
1+\frac{1-\alpha}{2}\frac{(2\ell^2+2\ell-1)}{(2\ell-1)(2\ell+3)}$
\end{tabular} \label{parametervalues}

\end{table}

\section{The interband susceptibility for C\lowercase{u}$_2$O}
\label{Cu2Osusceptibility} We consider the interband transition
between the highest valence band ($\Gamma^+_7$) and the lowest
conduction band ($\Gamma^+_6$) in Cu$_2$O. The conduction band and
the valence bands have the same parity and the dipole moment
between them vanishes. The $n\neq 1$ line corresponds to excitons
with the relative angular momentum $\ell=1$ and for this reason
the absorption process is dipole allowed. To compute the
susceptibility, we use the formula (\ref{susceptibility}). For the
sake of simplicity, we consider here the case of isotropic
effective electron and hole masses. The anisotropic case will be
considered below. The eigenfunctions $\varphi_n$ are the
hydrogen-like atom eigenfunctions
\begin{equation}
\varphi_n(r)~\to~\varphi_{n\ell m}(r)=R_{n\ell}(r)Y_{\ell
m}(\theta,\phi),
\end{equation}
 where
\begin{equation}\label{hydrogeneigenf}
R_{n\ell}(r)=C_{n\ell}\left(\frac{2r}{na^*}\right)^{\ell}M\left(-n+\ell+1,2\ell+2,\frac{2r}{na^*}\right)\exp\left(-\frac{r}{na^*}\right),
\end{equation}
 in terms of the Kummer function $M(a,b,z)$ with the normalization
\begin{equation}\label{normalizacja_atom_wodoru}
C_{n\ell}=\frac{1}{(2\ell+1)!}\left[\frac{(n+\ell)!}{2n(n-\ell-1)!}\right]^{1/2}\left(\frac{2}{na^*}\right)^{3/2},
\end{equation} and with the energies
\begin{equation}
E_n=-\frac{R^*}{n^2}, \qquad n=2,3,\ldots,
\end{equation}
$R^*$ being the effective excitonic Rydberg energy
\begin{equation}\label{erydberg}
R^*=\frac{\mu e^4}{2(4\pi\epsilon_0\epsilon_b)^2\hbar^2},
\end{equation}
and $\epsilon_b$ the bulk dielectric constant. We start with $P$
excitons and assume for ${\bf M}({\bf r})$ the form, appropriate
for the indirect gap (for the derivation, see Appendix
\ref{Appendix A}):
\begin{eqnarray}\label{gestoscwzbronione}
{\bf M}({\bf r})&=&
\textbf{e}_r\,M_{10}\frac{r+r_0}{2r^2r_0^2}e^{-r/r_0}=\textbf{e}_r
M(r) =\textbf{i}M_{10}\frac{r+r_0}{4{\rm
i}r^2r_0^2}\sqrt{\frac{8\pi}{3}}\left(Y_{1,-1}-Y_{1,1}\right)e^{-r/r_0}\nonumber\\
&&+\textbf{j}M_{10}\frac{r+r_0}{4r^2r_0^2}\sqrt{\frac{8\pi}{3}}\left(Y_{1,-1}+Y_{1,1}\right)e^{-r/r_0}+\textbf{k}M_{10}\frac{r+r_0}{2r^2r_0^2}
\sqrt{\frac{4\pi}{3}}Y_{10}e^{-r/r_0},
\end{eqnarray}
$r_0$ is the so-called coherence radius \cite{StB87}, \cite{CBT96}
\begin{equation}\label{r0}
r_{0}^{-1}=\sqrt{\frac{2\mu }{E_{g}}{\hbar^2}},
\end{equation}
$E_g$ the fundamental gap, and $\mu$ the reduced effective mass
for the pair electron-hole. The above expression gives the
coherence radius in terms of
 effective band parameters, but we find it
convenient to treat the coherence radii as free parameters which
can be determined by fitting experimental spectra.  Mostly one
takes it as a fraction of the respective excitonic Bohr radius.
Taking into account the $\ell=1$ states and the
\textbf{k}-component of the dipole density
(\ref{gestoscwzbronione}) and restricting the consideration to the
resonant states, we obtain from (\ref{susceptibilitychi}) the
$P$-exciton counterpart of the susceptibility

\begin{eqnarray}
\chi_P(\omega,\textbf{k})&=&\frac{2}{\epsilon_0\hbar}\sum\limits_{n=2}^N\frac{b_{n1}}{\Omega_n(\textbf{k})-(\omega+{\rm
i}\gamma_n(k))},\\
\label{coefficients1}b_{n1}&=&\frac{8\pi}{3}\left(\int\limits_0^\infty
{\rm d}r r^2 R_{n1}(r)M(r)\right)^2,
\end{eqnarray}
 $R_{n1}$ are the $p-$ symmetric radial hydrogen atom
eigenfunctions, see (\ref{hydrogeneigenf}) for $\ell=1$
\begin{equation}\label{wodorradialn1}
R_{n1}(r)=C_{n1}\left(\frac{2r}{na^*}\right)M\left(-n+2,4,\frac{2r}{a^*n}\right)\exp\left(-\frac{r}{na^*}\right),
\end{equation}
with the normalization
\begin{equation}\label{normalizacja_atom_wodoru1}
C_{n1}=\frac{1}{3!}\left[\frac{(n+1)!}{2n(n-2)!}\right]^{1/2}\left(\frac{2}{na^*}\right)^{3/2},
\end{equation}
where the excitonic Bohr radius $a^*$ is used
\begin{equation}\label{ebohr}
a^*=\frac{(4\pi\epsilon_0\epsilon_b)\hbar^2}{\mu e^2}.
\end{equation}
After performing the calculations (see Appendix \ref{Appendix B})
we obtain the coefficients $b_{n1}$ in the form
\begin{eqnarray}\label{coefficientsbn1}
&&b_{n1}=M_{10}^2\frac{n^2-1}{n^5}\frac{16\pi}{3}\left(\frac{a^*}{r_0}\right)^4\frac{1}{a^{*3}}\left(\frac{nr_0}{r_0+na^*}\right)^6.
\end{eqnarray}

\section{Polariton dispersion
relation}\label{polariton.dispersion} 
\begin{figure}[ht!]
\centering
\includegraphics[width=0.6\linewidth]{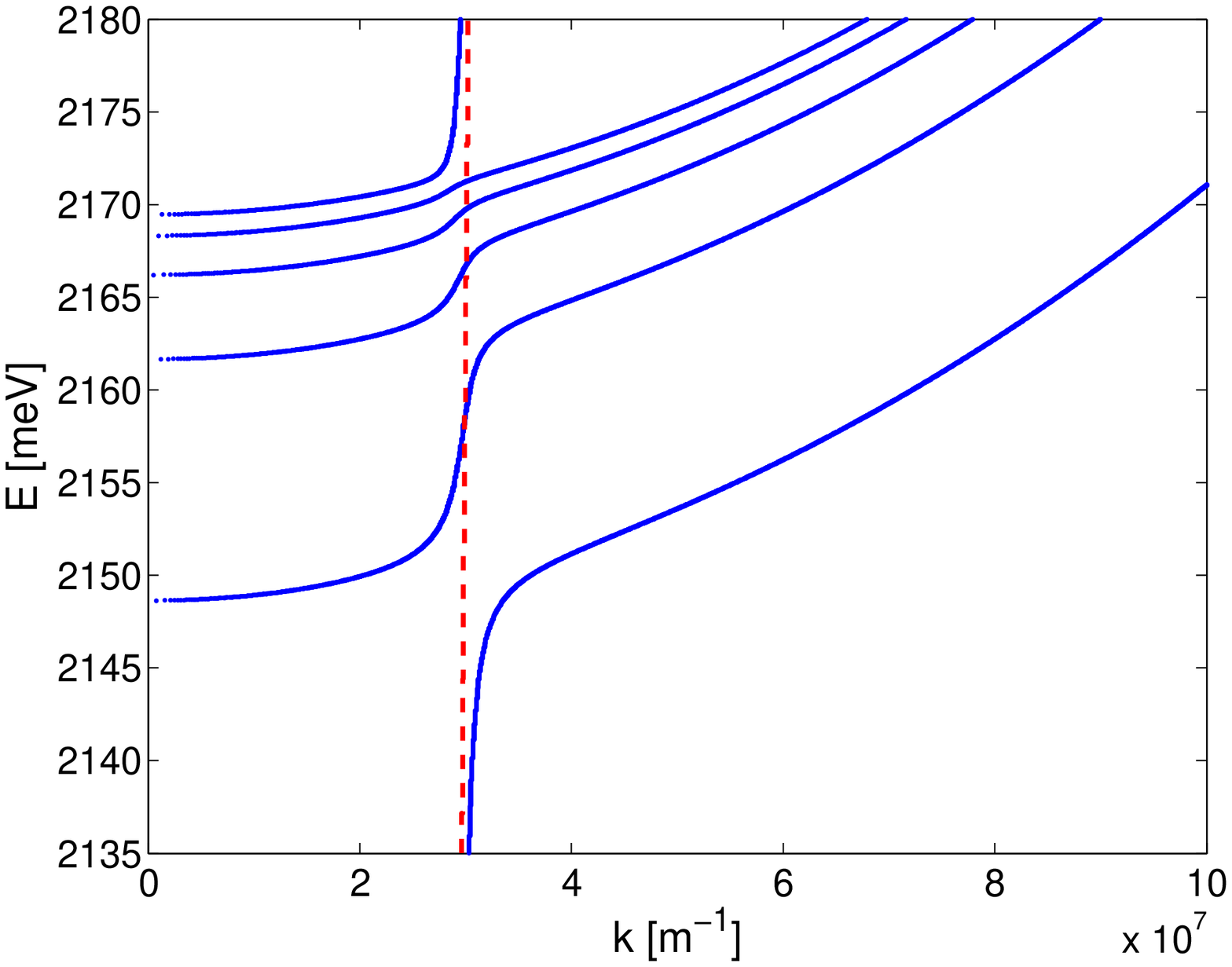}
  \caption{\small  The polariton dispersion for a Cu$_2$O crystal, calculated by eq. (\ref{polariton1}), taking into account the 5 lowest excitonic states.}
\label{FigDisp}
\end{figure}

Having the coefficients
$b_{n1}$ and thus the $P$-exciton part of the susceptibility, we
obtain from (\ref{Maxwell}) the polariton dispersion relation
\begin{equation}\label{polariton}
\frac{c^2k^2}{\omega^2}-\epsilon_b=\frac{2}{\epsilon_0\hbar}\sum\limits_{n=2}^N\frac{b_{n1}}{\Omega_n(\textbf{k})-(\omega+{\rm
i}\gamma_n(k))}.
\end{equation}
It follows that for $k=0$ and $\gamma_n=0$ the r.h.s. exhibit
resonances located at the transversal frequencies $\omega_{Tn}$.
In the case under consideration
\begin{equation}
\hbar\omega_{Tn}=E_{Tn}=\hbar\omega_g-\frac{R^*}{n^2},\qquad
n=2,3,\ldots.
\end{equation}
Using the oscillator strengths $f_{n1}$ defined in
Appendix~\ref{Appendix B}, the polariton dispersion relation takes
the form
\begin{equation}\label{polariton1}
\frac{k^2}{k_0^2}-\epsilon_b=\epsilon_b\sum\limits_{n=2}^N\frac{f_{n1}\Delta^{(2)}_{LT}/R^*}{\left(E_{Tn}-E-{\rm
i}{\mit\Gamma}\right)/R^* +(\mu/M)(ka^*)^2}
\end{equation}
where $k_0=\omega/c$. The resulting polariton dispersion shape is
displayed in Fig. \ref{FigDisp}.
In the above dispersion relation only $P$ excitons are considered.
The relation will much more complicated when excitons with higher
angular momentu number will be included, for example $F$ or $H$
excitons.
\section{Anisotropy effects}\label{anisotropy}
The main difference between the effective hole and electron masses
in Cu$_2$O is the high anisotropy observed for the hole masses
while, as expected by symmetry, the electron value remains
practically the same for all directions. The anisotropy in the
hole effective masses is evidenced by their very different
components in the [100], [110], and [111] directions (see, for
example, Refs.~\cite{Ching}, \cite{Ruiz}). As was shown by Dasbach
et al.~\cite{Dasbach} (see also~\cite{Froelich}), the total
exciton mass in the $[0,0,1]$ direction equals $3~m_0$, whereas
the value in the $[1,1,0]$ direction is equal 0.66 (for
example,~\cite{Kazimierczuk}) or 0.5587 (\cite{Thewes}), see
Table~\ref{parametervalues}. The effective mass anisotropy plays
an important role in the description of the optical properties of
excitons. In the RDMA the anisotropy paramer $\alpha$ is defined,
$\alpha=\mu_\parallel/\mu_z$, $\mu_\parallel=\mu[110]$ and $\mu_z$
being the electron-hole reduced masses in the respective
directions. This parameter is then used in modified expressions
for the excitonic eigenfunctions and eigenvalues. Instead of
(\ref{hydrogeneigenf}) we will use
\begin{equation}\label{wodorradial}
R_{n\ell}(r)=C_{\ell}\left(2\lambda r\right)^{\ell}M\left(\ell +1
-\frac{\eta_{\ell m}}{\lambda},2\ell+2,2\lambda
r\right)\exp\left(-\lambda r\right),
\end{equation}
with $\lambda=\sqrt{-E}$, $C_\lambda$ is the normalization factor,
and
\begin{equation}\label{eta}
\eta_{\ell m}(\alpha)=\int\limits_0^{2\pi}{\rm
d}\phi\int\limits_0^\pi\frac{\vert Y_{\ell m}\vert^2
\sin\theta\,{\rm
d}\theta}{\sqrt{\sin^2\theta+\alpha\cos^2\theta}}.
\end{equation}
 Bound states appear when the index $(\ell +1-\eta_{\ell
 m}/\lambda)$ attains zero or a negative integer, so that the
 discrete eigenvalues are given by \cite{RivistaGC},
 \cite{BCT95a}:
\begin{equation}
E_{n\ell m}=-\frac{\eta^2_{\ell m}(\alpha)R^*}{n^2},\quad
n=1,2,\ldots,\quad \ell=0,1,2,\ldots,n-1,\quad
m=0,1,2,\ldots,\ell.
\end{equation}
For the lowest eigenvalues we can use another expressions
\begin{equation}\label{aenel}
E_{n\ell}=-A_{n\ell}^2R^*,
\end{equation}
with
\begin{eqnarray}
A_{10}&=&\frac{2}{1+\sqrt{\alpha}},\nonumber\\
A_{20}&=&\frac{2(1+2\sqrt{\alpha})}{3(1+\sqrt{\alpha})^2},\\
A_{30}&=&\frac{1}{15}\biggl\{\biggl[\frac{2}{1+\sqrt{\alpha}}\nonumber\\
&&\times\left(\frac{\sqrt{\alpha}}{1+\sqrt{\alpha}}+2\frac{\sqrt{\alpha}}{(1+\sqrt{\alpha})^2}+\sqrt{\alpha}+3\right)\biggr]\biggr\}\nonumber
\end{eqnarray}
etc., see Ref. \cite{BCT95a}. The above formulas can explain the
huge, as compared to the Rydberg energy, excitonic binding energy,
for which we obtained the values 108.95 meV (105.32 meV),
depending on the parametrs  used, see Table~\ref{parametervalues}.
This values show the property of  the observed exciton binding
energy in Cu$_2$O, which is greater than the corresponding Rydberg
energy (see, for example, \cite{Kavoulakis} and, more recently,
\cite{Frazer}). We can also calculate the $P$ and $F$ excitons
resonances, using the values of the parameters $\eta_{10}$ and
$\eta_{30}$. The results show that the $F$ exciton resonances are
shifted to the higer energy, compared to the $P$ excitons (see
Table~\ref{parametervalues}), as was described in
Ref.~\cite{Thewes}. Using the above results we can extend the
expressions for the susceptibility, including the effects of $F$
excitons

\begin{eqnarray}\label{susceptiblityanisotropic}
\chi&=&\epsilon_b\sum\limits_{n=2}^N\frac{f_{n1}\Delta^{(2)}_{LT}/R^*}{\left(E_{Tn10}
-E-{\rm i}{\mit\Gamma}\right)/R^*+(\mu_\parallel/M)(ka^*)^2}\nonumber\\
&&+\epsilon_b\sum\limits_{n=4}^N\frac{f_{n3}\Delta^{(2)}_{LT}/R^*}{\left(E_{Tn30}
-E-{\rm i}{\mit\Gamma}\right)/R^* +(\mu_\parallel/M)(ka^*)^2},
\end{eqnarray}
where the excitonic resonances are given at the energies
\begin{eqnarray}
E_{Tn10}&=&-\frac{\eta_{10}^2}{n^2}R^*,\quad n=2,3,\ldots,\qquad
E_{Tn30}=-\frac{\eta_{30}}{n^2}R^*,\quad n=4,5,\ldots
\end{eqnarray}
and
\begin{equation}
f_{n3}=\frac{(n^2-9)(n^2-4)(n^2-1)}{n^9}\cdot f,
\end{equation}
where we set all remaining constants, including the unknown dipole
matrix element $M_{03}$ as the factor $f$, which can be obtained,
for example, by fitting the experimental spectra.
 In particular, the resonances for the $n=4$ state will be
observed at the energies $E_{T410}=E_g-E_{410}=2158.6~\hbox{meV}$
($P$ exciton) and $E_{T430}=E_g-E_{430}=2160~\hbox{meV}$ ($F$
exciton). This splitting is qualitatively in agreement with the
observation by Thewes \emph{et al}~\cite{Thewes}. The differences
can be explained by the fact, that in Ref.~\cite{Thewes} another
set of parameters (Rydberg energy, effective masses) was used, as
can be seen in Table~\ref{parametervalues}. Using the data of
Ref.~\cite{Thewes} we obtain practically the same values for the
excitonic resonances.
\section{Green's function method}\label{Green}
The constitutive equation (\ref{constitutiveeqn}) is a
nonhomogeneous differential equation, which can be solved by using
the appropriate Green's function. For bulk crystals, assuming the
harmonic time-space dependence and only one valence$\to$conduction
band transition, the Green function of the l.h.s. operator in
(\ref{constitutiveeqn}) satisfies the equation

\begin{equation}\label{Greenogolnyzpotencjalem}
\left(E_g-\hbar\omega-{\rm i}{\mit\Gamma}+\frac{\hbar^2
k^2}{2M}-\frac{\hbar^2}{2\mu}\hbox{\boldmath$\nabla$}^2-\frac{e^2}{4\pi\epsilon_0\epsilon_b
r}\right)G\left({\bf r},{\bf r}'\right)=\delta\left({\bf r}-{\bf
r}'\right).
\end{equation}
\noindent Using the expansion
\begin{equation}
\delta\left({\bf r}-{\bf
r}'\right)=\frac{\delta\left(r-r'\right)}{r^2}\sum\limits_{\ell=0}^{\infty}\sum\limits_{m=-\ell}^{\ell}Y^*_{\ell
m}\left(\theta',\phi'\right)Y_{\ell m}(\theta,\phi),
\end{equation}
\noindent where $Y_{\ell m}(\theta,\phi)$ are spherical harmonics,
we look for the Green function in the form
\begin{equation}\label{rozwiniecieharmoniki}
G\left({\bf r}, {\bf
r}'\right)=\sum\limits_{\ell=0}^{\infty}\sum\limits_{m=-\ell}^{\ell}Y^*_{\ell
m}(\theta',\phi')Y_{\ell m}(\theta,\phi)g_{\ell m}(r,r').
\end{equation}
\noindent Functions $g_{\ell m}$ satisfy the equations

\begin{equation}\label{glmcoulomb}
\left(\frac{{\rm d}^2}{{\rm d}r^2}+\frac{2}{r}\frac{{\rm d}}{{\rm
d}r}+\frac{2}{r}-\frac{\ell(\ell+1)}{r^2}-\kappa^2\right)g_{\ell
m}=-\frac{2\mu}{\hbar^2a^*}\frac{\delta\left(r-r'\right)}{r^2},
\end{equation}
\noindent where $r$ is scaled in the excitonic Bohr radii
(\ref{ebohr}), and
\begin{equation}\label{kappaskalowane}
\kappa^2=\frac{2\mu}{\hbar^2}a^{*2}\left(E_g-\hbar\omega-{\rm
i}{\mit\Gamma}\right)+\frac{\mu}{M}\left(ka^*\right)^2.
\end{equation}
\noindent The solution of (\ref{glmcoulomb}) is given in the form
(for example, \cite{StB87})
\begin{equation}\label{greencoulomb}
g_{\ell m}\left(r,r'\right)=C\left(4\kappa^2rr'\right)^{\ell}
e^{-\kappa\left(r+r'\right)}M\left(a_{\ell},b_{\ell},2\kappa
r^<\right)U\left(a_{\ell},b_{\ell},2\kappa r^>\right),
\end{equation}
\noindent where  $r^<=$ min $(r,r')$, $r^>$= max $(r,r')$,
\begin{eqnarray}\label{C}
C&=&\frac{2\mu}{\hbar^2a^*}\frac{2\kappa \Gamma
\left(a_{\ell}\right)}{\Gamma \left(b_{\ell}\right)}, \qquad
a_{\ell}= \ell +1-\frac{1}{\kappa},\qquad b_{\ell}= 2\ell+2,
\end{eqnarray}
 \noindent
$M(a,b,c)$, $U(a,b,c)$ are $Kummer$ functions (confluent
hypergeometric functions) in the notation of Abramowitz {et
al.}~\cite{Abramowitz},  $Y_{\ell m}$ are spherical harmonics,
$\Gamma(z)$ is the Euler Gamma-function,
 $a^*$ is the effective excitonic Bohr radius (\ref{ebohr}). Having the Green function, we calculate the coherent
amplitudes
\begin{equation}\label{YGreen}
Y({\bf R},{\bf r})= {\bf E}({\bf R})\int {\rm d}^3{ r}' {\bf  M}
({\bf r}')G({\bf r},{\bf r}'),
\end{equation}
\noindent and thus the excitonic polarization from
eq.~(\ref{polarization})
\begin{equation}\label{PolarizationGreen}
{\bf P}({\bf R})=\,2\int\int {\rm d}^3 r {\rm d}^3r'{\bf M} ({\bf
r}) G({\bf r},{\bf r}'){\bf M}({\bf r}')\cdot{\bf E}({\bf R}).
\end{equation}
\noindent The linear dielectric susceptibility tensor
\begin{equation}\label{MGM}
\epsilon_0\underline{\underline{\chi}}(\omega,k)=2\int\int {\rm
d}^3 r {\rm d}^3 r'{\bf M} ({\bf r}) G({\bf r},{\bf r}'){\bf
M}({\bf r}')
\end{equation}
\noindent relates the electric field vector ${\bf E}$ to the
polarization vector ${\bf P}$, and specific polarization
components can be considered.
 From the above Green function expressions
(\ref{YGreen}), (\ref{PolarizationGreen}) and (\ref{MGM}) can be
computed once an appropriate expression for ${\bf M}({\bf r})$ is
considered.

Below the gap the imaginary part of the susceptibility
$\chi(\omega,k)$  shows maxima in correspondence to  the excitonic
energy resonances. They are obtained from the singularities in the
Green functions $g_{\ell m}$ and for $k=0, \Gamma=0$ are given by:

\begin{eqnarray}
E_{Tn}&=&E_g+E_n,\nonumber\\
E_{n}&=&-\frac{R^*}{n^2},\qquad n=1,2,\ldots. \label{eigen1}
\end{eqnarray}
To simplify the calculations, we will use the transition dipole
intensity in the form \cite{StB87}
\begin{equation}
\textbf{M}(\textbf{r})=\textbf{M}_{10}\frac{\textbf{r}}{r_0^3}\delta(r-r_0),
\end{equation}
 and compute the
susceptibility tensor element $\chi_{zz}$
\begin{eqnarray}
\chi_{zz}(\omega,k)&=&\frac{2}{\epsilon_0}\int\int {\rm d}^3 r
{\rm d}^3 r'M_z ({\bf r}) G({\bf r},{\bf r}')M_z({\bf
r}')\nonumber\\&=&\frac{2}{\epsilon_0}\int\int {\rm d}^3 r {\rm
d}^3 r'M_z ({\bf
r})\sum\limits_{\ell=0}^{\infty}\sum\limits_{m=-\ell}^{\ell}Y^*_{\ell
m}(\theta',\phi')Y_{\ell m}(\theta,\phi)g_{\ell m}(r,r')M_z({\bf
r}')\\
&=&\frac{4\pi^2}{3}\cdot
\frac{2\mu}{\hbar^2}\epsilon_b\frac{2M_{10}^2}{\epsilon_0\epsilon_b\pi
a^*}\cdot
g_{10}(r_0,r_0)=\frac{4\pi^2}{3}\cdot\epsilon_b\frac{\Delta_{LT}}{R^*}g_{10}(r_0,r_0),\nonumber
\end{eqnarray}
with
\begin{eqnarray}
g_{10}(r_0,r_0)&=&\frac{\kappa \Gamma
\left(a_{1}\right)}{3}\left(4\kappa^2r_0^2\right)e^{-2\kappa
r_0}M\left(a_{1},4,2\kappa
r_0\right)U\left(a_{1},4,2\kappa r_0\right),\nonumber\\
&&a_{1}= 2-\frac{1}{\kappa},\qquad b_{1}= 4,
\end{eqnarray}
where $r_0$ is expressed in the units of $a^*$, and
(\ref{kappaskalowane}) is used in the form
\begin{equation}\label{kappaskalowane1}
\kappa^2=\frac{E_g-\hbar\omega-{\rm
i}{\mit\Gamma}}{R^*}+\frac{\mu}{M}\left(ka^*\right)^2.
\end{equation}
In a similar way one can calculate contributions of higher order excitonic states (F,H...)
Having the susceptibility tensor elements obtained in terms of the
Green function, we can alternatively compute the bulk absorption
(in the limit $k\to 0$) and the polariton dispersion.

\section{Results of specific calculations}\label{results}
\begin{figure}[h]
\includegraphics[width=0.5\linewidth]{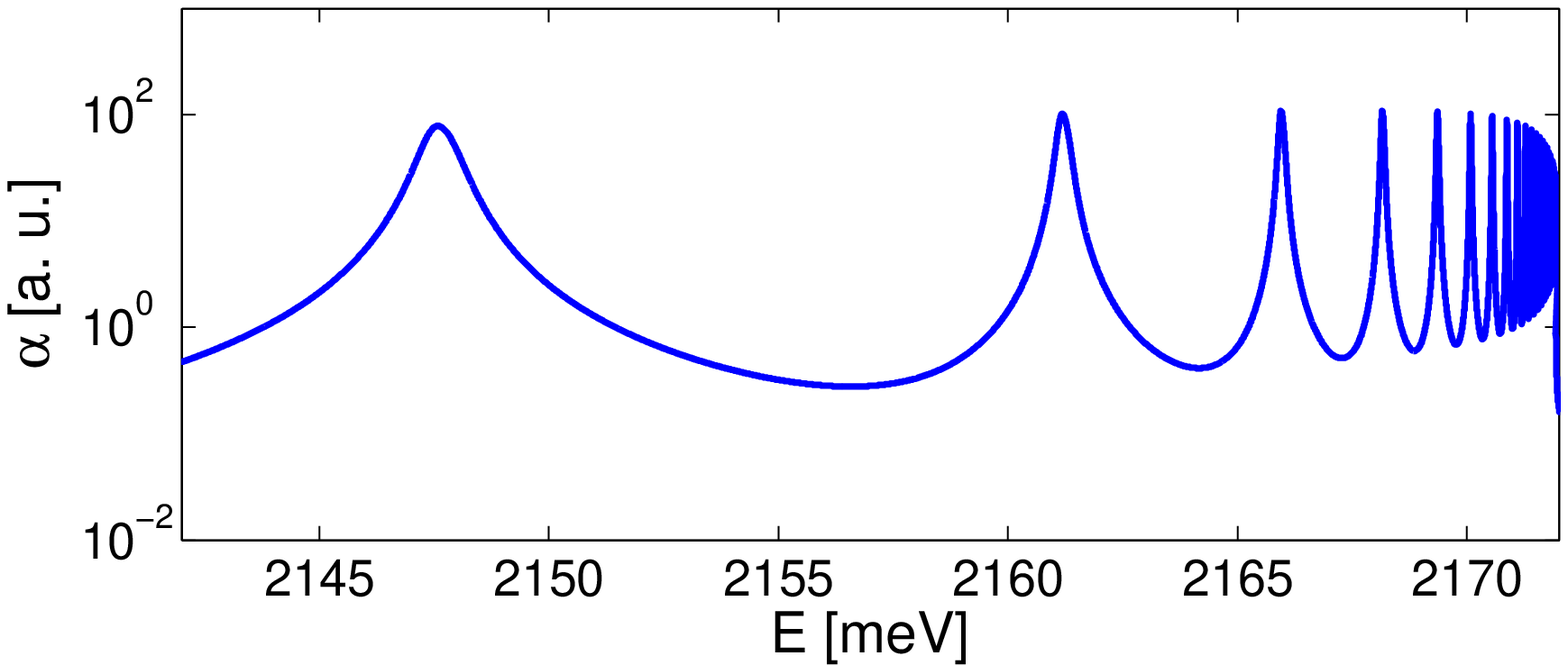}
\hfill
\includegraphics[width=0.5\linewidth]{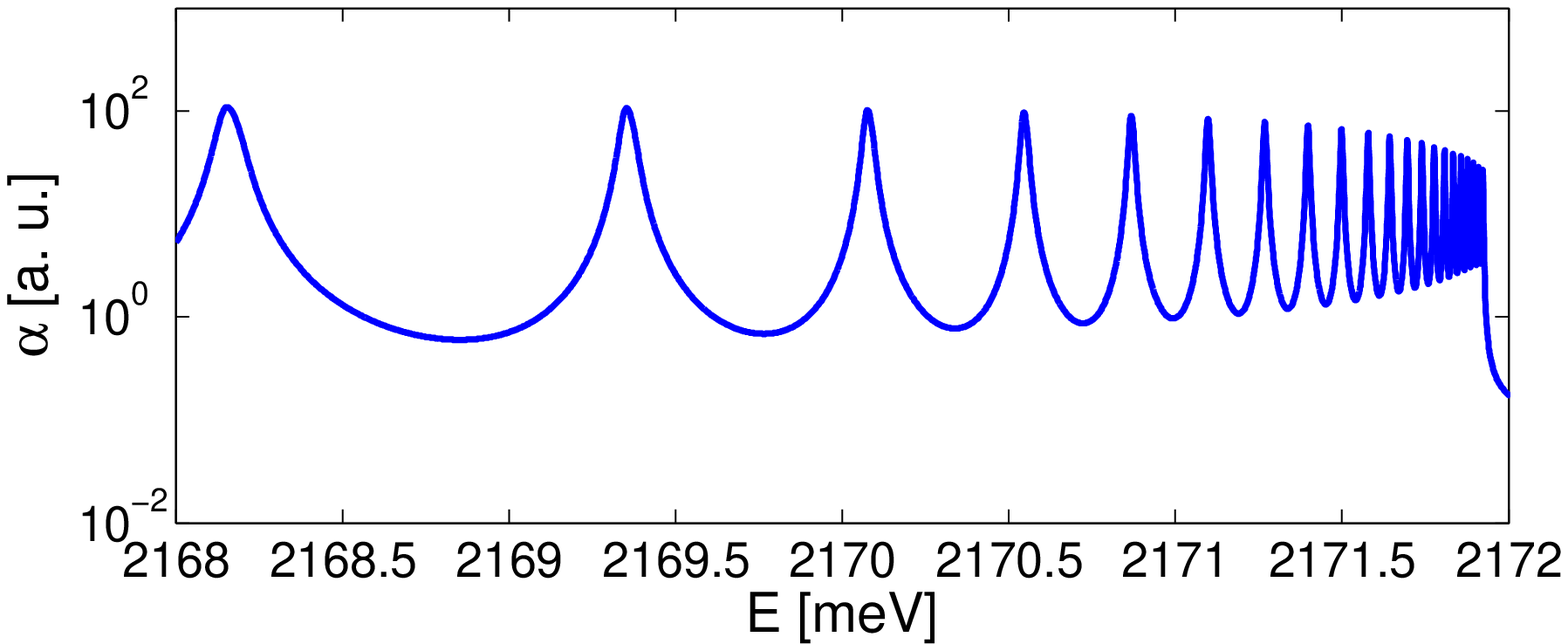}
\hfill
\includegraphics[width=0.5\linewidth]{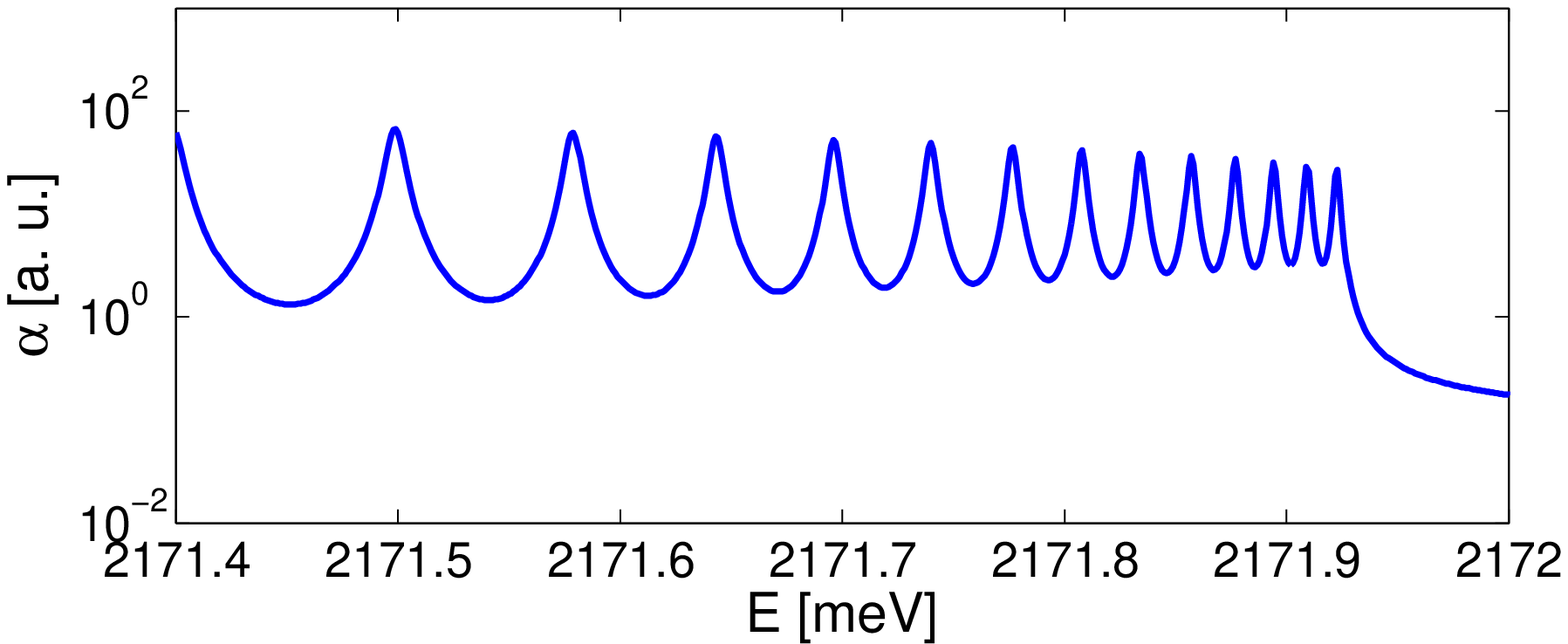}
\includegraphics[width=0.5\linewidth]{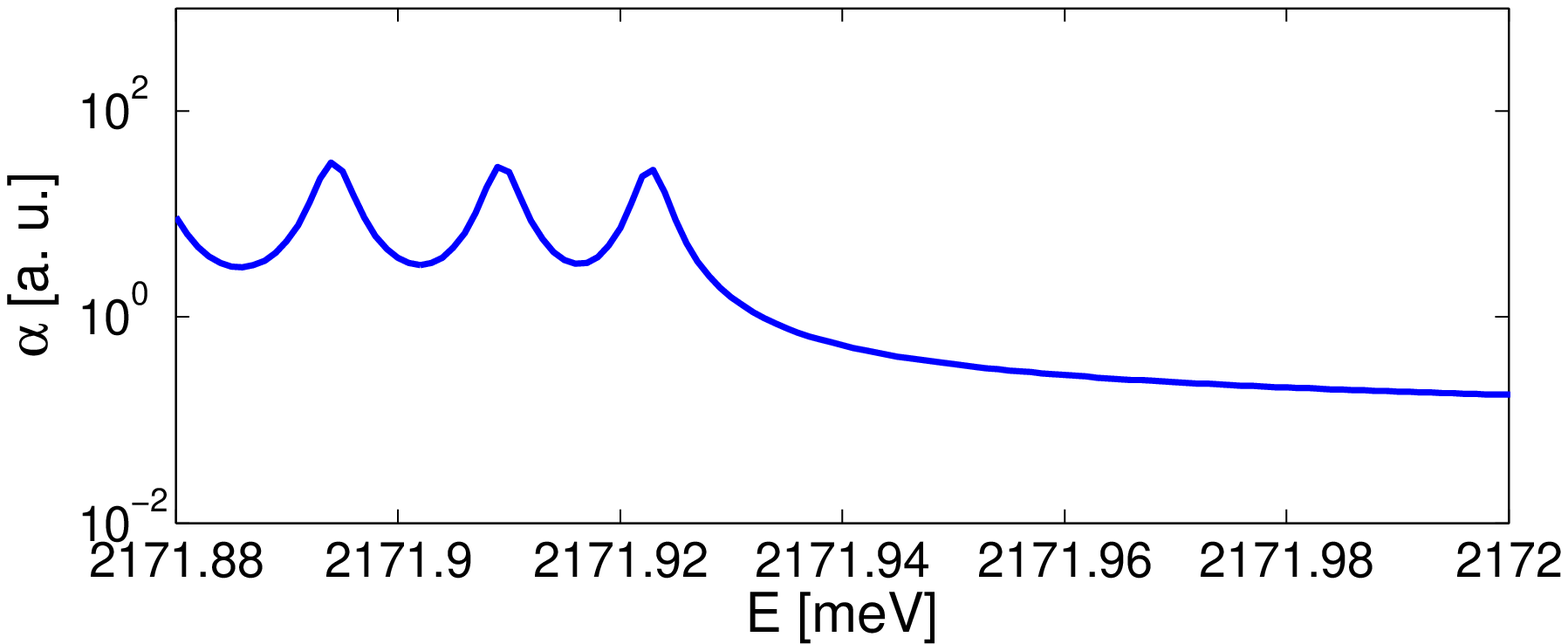}
\hfill
  \caption{\small The bulk absorption of a Cu$_2$O crystal
calculated by eq. (\ref{absorption}) for four energy intervals
below the fundamental gap. The logarithmic scale is applied.}
\label{Fig1ab}
\end{figure}

\begin{figure}[ht!]
\centering
\includegraphics[width=0.45\linewidth]{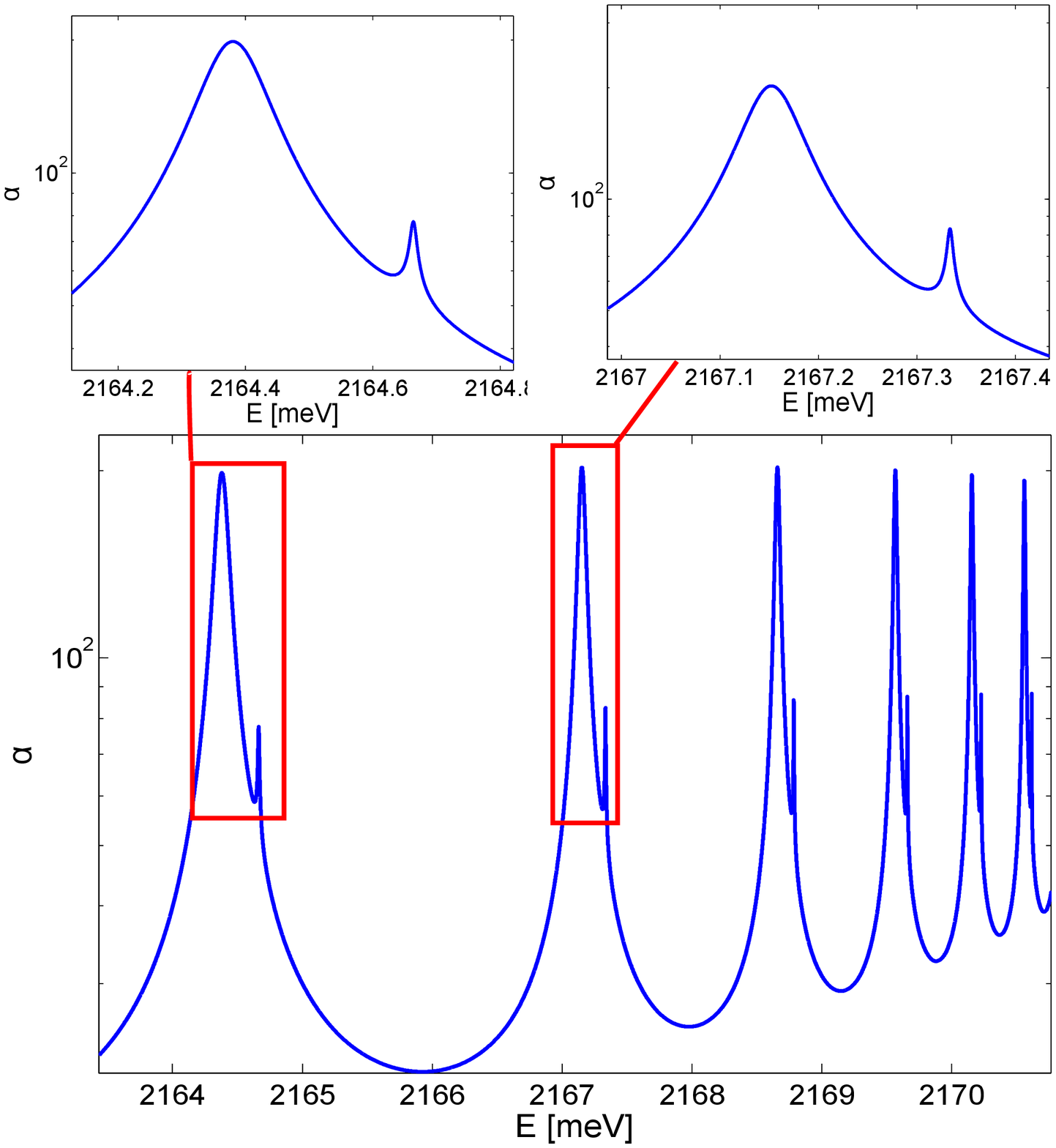}
  \caption{\small  The absorption spectra including the effect of $P$ and $F$ excitons, calculated using the
   parameters from Ref.~\cite{Thewes}, see
   Table~\ref{parametervalues}}
\label{FigThewes}
\end{figure}

We performed calculations for the Cu$_2$O crystal having in mind
the experiments by Kazimierczuk et al.~\cite{Kazimierczuk}. The
 parameters  used in calculations are collected in Table~\ref{parametervalues}, where the  Rydberg energy $95.74~\hbox{meV}$ and the effective excitonic Bohr radius $a^*=1.00~\hbox{nm}$ result from \ref{erydberg} and \ref{ebohr}.
First we calculated the polariton dispersion relation, shown in
Fig. \ref{FigDisp}. We observe a multiplicity
of values of the wave vectors, which means a multiplicity of
polariton waves, much greater than those observed in other
semiconductors as, for example, GaAs. This fact should certainly
be regarded by computing the optical functions such as, for
example, reflectivity, where the polaritonic aspect is relevant.
Than we computed the absorption, shown in Fig. \ref{Fig1ab}. Since
the absorption peaks decrease quite rapidly, we applied the
logarithmic scale. When the excitonic contributions of $F$ and $H$
are included, we observe the occurrence of additional absorption
peaks, as shown in Fig.~\ref{FigThewes}.

\begin{figure}[h]
\includegraphics[width=0.35\linewidth]{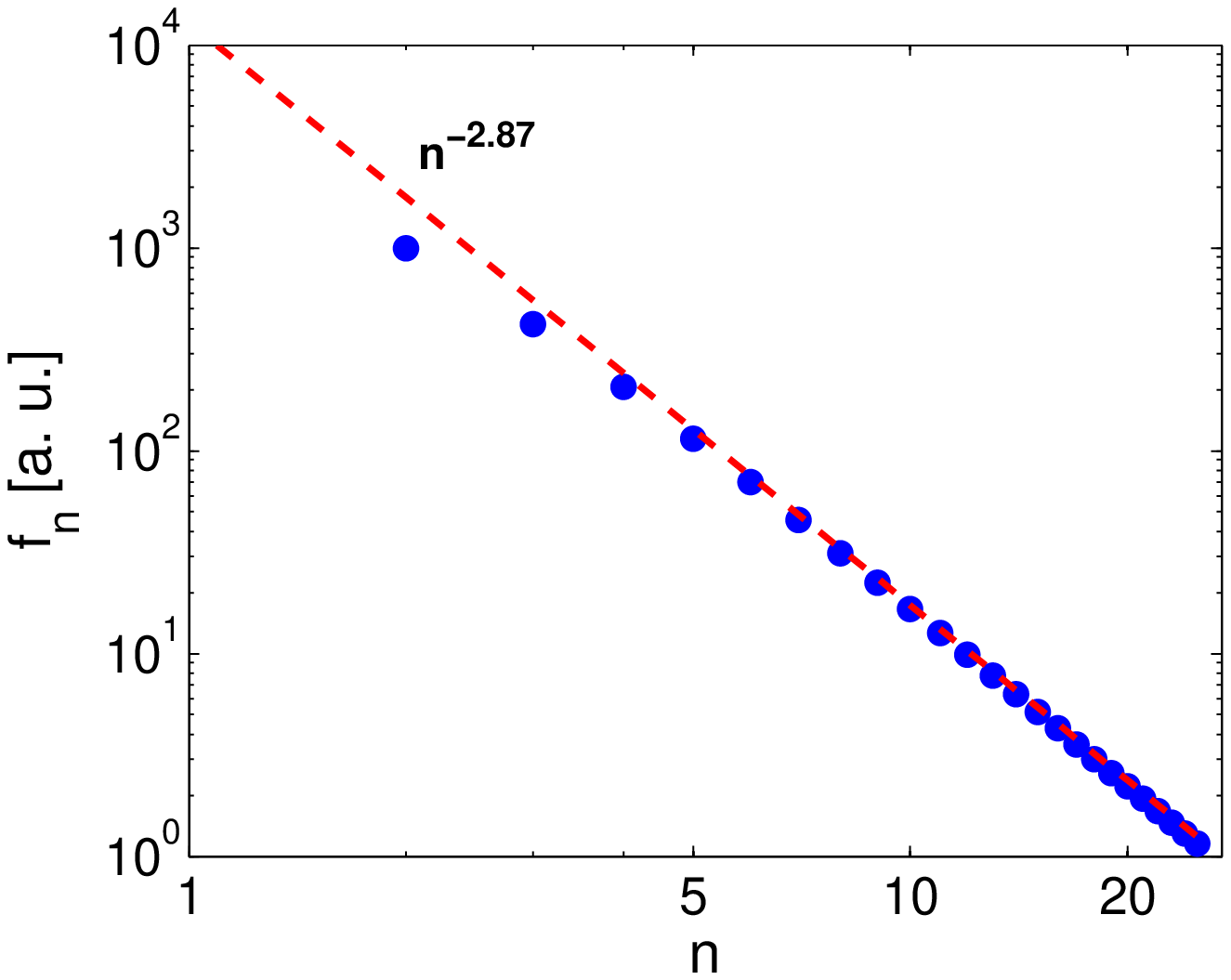} a)
\hfill
\includegraphics[width=0.35\linewidth]{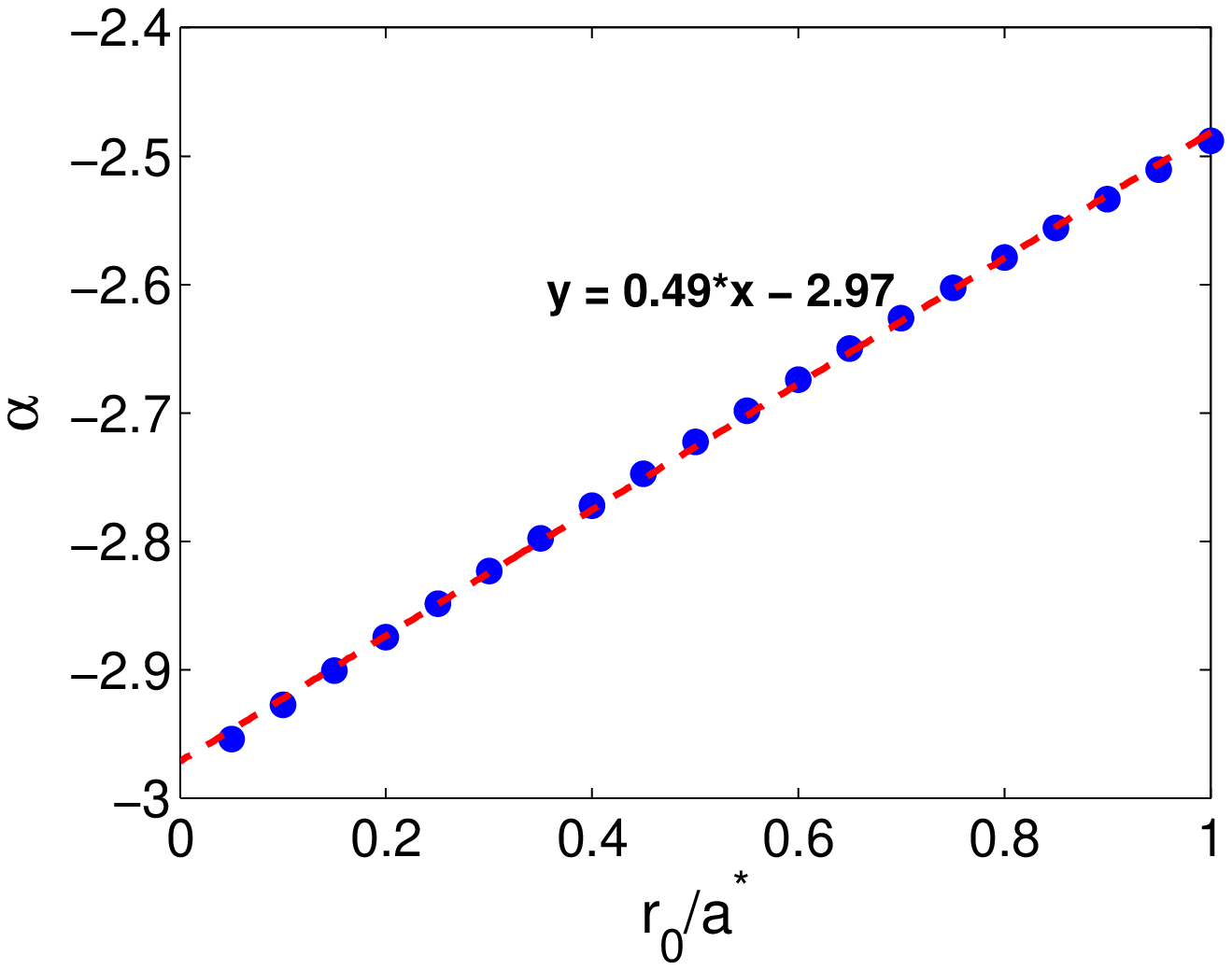} b)
  \caption{\small a) The dependence of the oscillator forces $f_n$ on the state
number $n$ with a linear fit added, b) The exponent $\alpha$ in the $n^{-\alpha}$ law as a function of the coherence radius $r_0$}
\label{Fig2ab}
\end{figure}

\begin{figure}[h]
\includegraphics[width=0.35\linewidth]{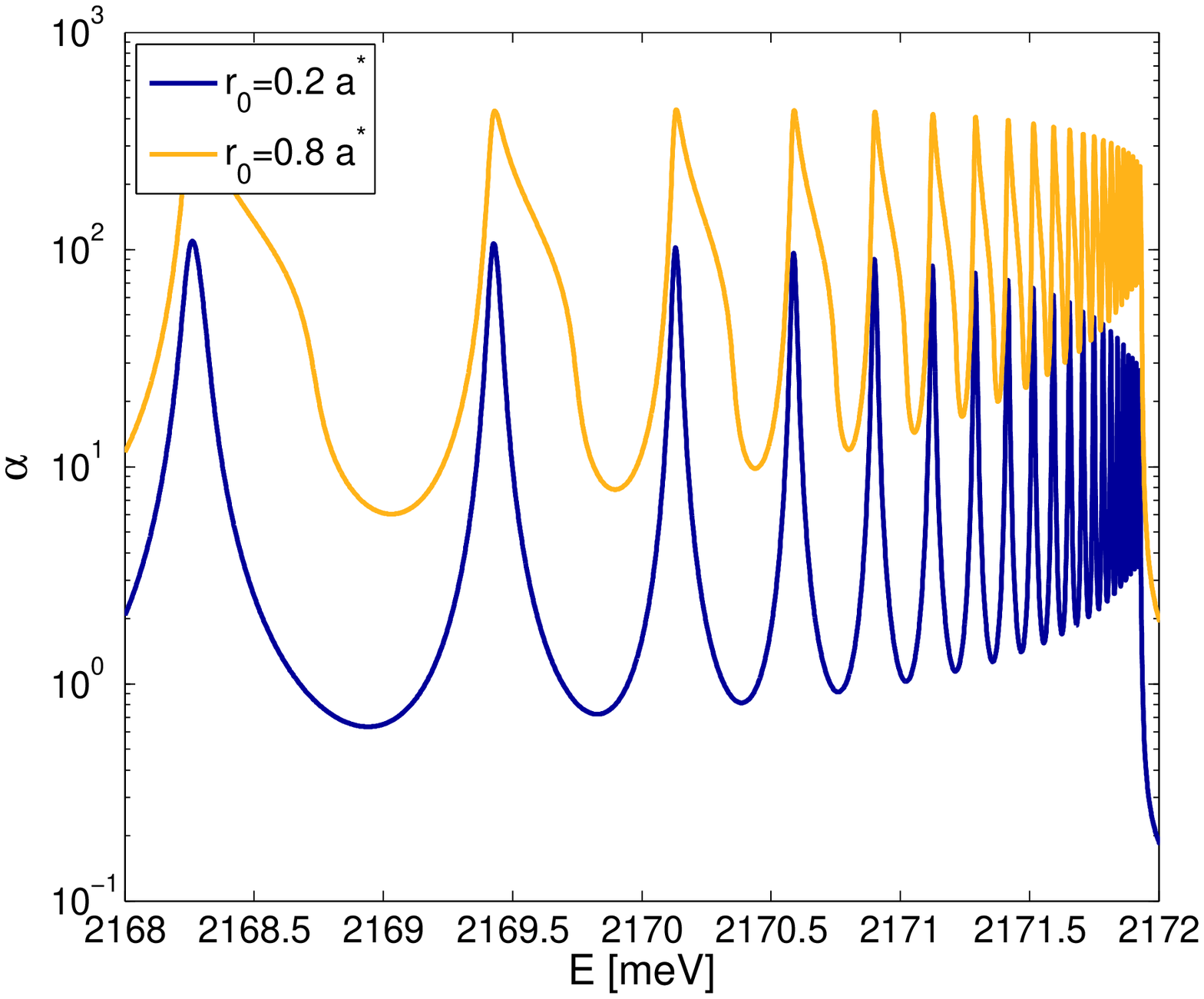}
\hfill
\includegraphics[width=0.35\linewidth]{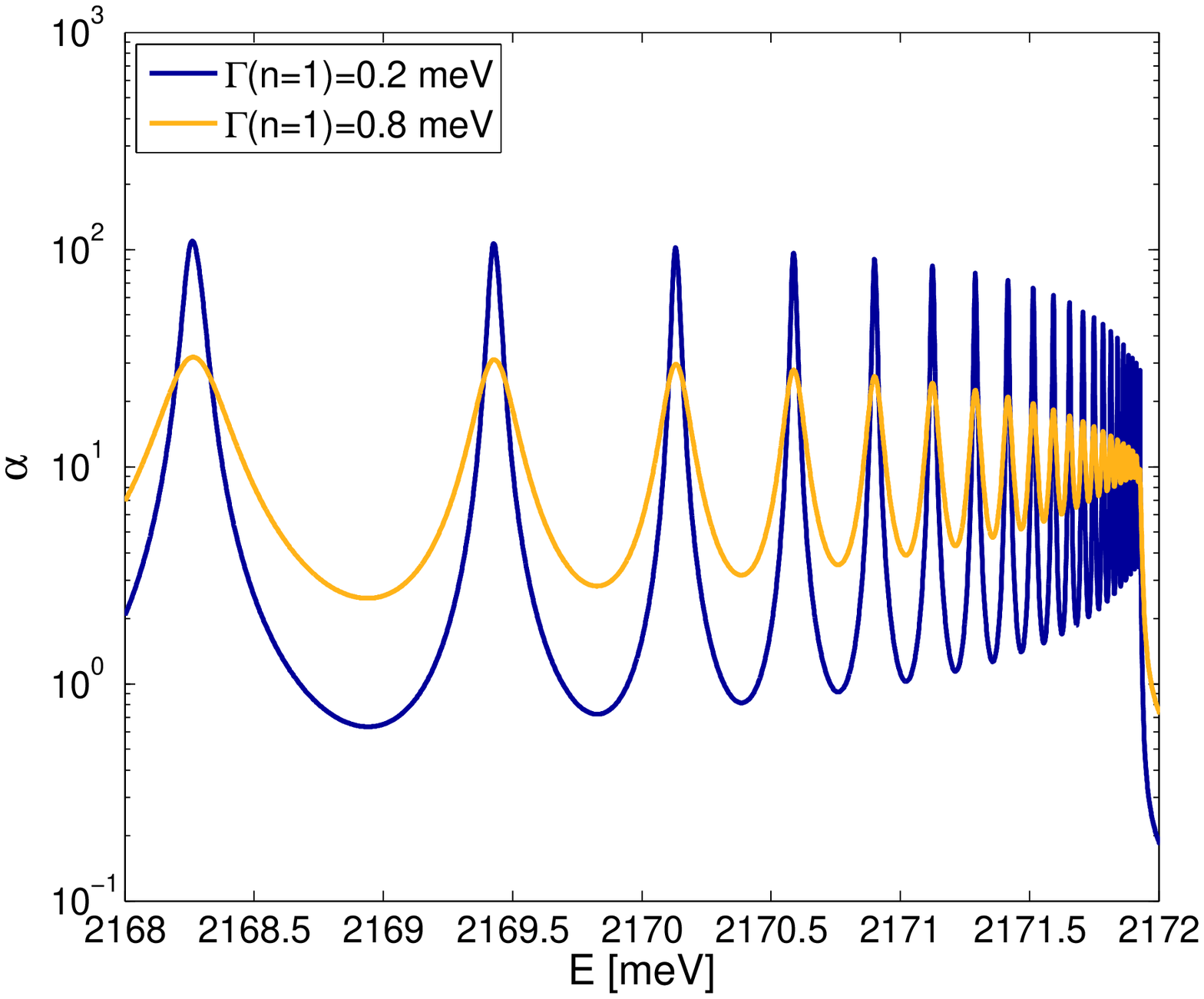}
  \caption{\small  a) The dependence of the absoption line shape, calculated for two values of the coherence radius $r_0$, b) The dependence on the damping constant $\Gamma$}
\label{Fig3}
\end{figure}

By the relation (\ref{oscillatorforces}) we can establish the
dependence between the oscillator forces $f_{n1}$ and the
corresponding state number $n$. The oscillator forces decrease
with the number $n$, but the slope also depends on the coherence
radius $r_0$. The general dependence is shown in Fig.
\ref{Fig2ab}, and in Fig. \ref{Fig3} we display the dependence on
the coherence radius $r_0$. As it can be seen, the relation
$f_n\propto n^{-3}$ is obtained only in the limit $r_0\to 0$. The
best fit to the experimental spectra is obtained for $n^{-2.87} $,
as shown in Fig.~\ref{Fig2ab}

In Fig. \ref{Fig3} we present the dependence of the absorption
line shape on the magnitude of the coherence radius $r_0$.
In the cases of semiconductors with large excitonic Bohr radii
$a^*$ (large compared to the lattice constant as, for example, in
GaAs) the coherence radius was taken as a small fraction of $a^*$.
In the case of Cu$_2$O the Bohr radius is not very large (about
twice) compared to the lattice constant. Therefore also $r_0$
cannot be very small. We can take, for example $r_0=0.5 a^*$ but
the exact value will be estimated by fitting the experimental
spectra. As reported, for example, in~\cite{Klingshirn}, the
longitudinal-transversal energy in Cu$_2$O is of the order of a
few $\mu\hbox{eV}$. We put $\Delta_{LT}^{(2)}=10\mu\hbox{eV}$.
Using eq. (\ref{polariton1}) and taking into account the lowest 25
excitonic states, we can determine the absorption coefficient from
the relation
\begin{equation}\label{absorption}
\alpha=10^2\frac{E\;[\hbox{meV}]} {1.976~\hbox{meV}\cdot~\hbox
{cm}}\sqrt{\epsilon_b}~\hbox{Im}~\left[1+\sum\limits_{n=2}^{25}\frac{f_{n1}\Delta^{(2)}_{LT}}{E_{Tn10}-E-{\rm
i}{\mit\Gamma}_{n1}}+\sum\limits_{n=4}^{25}\frac{f_{n3}\Delta^{(2)}_{LT}}{E_{Tn30}-E-{\rm
i}{\mit\Gamma}_{n3}}+\ldots\right]^{1/2},
\end{equation}
where only $m=0$ states where accounted for.

Recently, we became aware of a new experiment concerning absorption spectra which
have taken into account the probability for absorbing photons from a laser as well as
contributions relating to both coherent and incoherent parts of the spectrum of the excitons by Gr\"unwald et al \cite{Grunwald}.
\section{Final remarks}\label{finalremarks}
The main results of our paper can be summarized as follows. We
have proposed a procedure based on the RDMA approach that allows
to obtain analytical expressions for the optical functions of
semiconductor crystals including high number Rydberg excitons,
also for the case of indirect interband transitions. Our results
have general character because arbitrary exciton angular momentum
number is included.  The effect of anisotropic dispersion and
coherence of the electron and the hole with the radiation field is
outlined. The theoretical findings are confirmed by numerical
simulations. We have chosen the example of cuprous dioxide,
inspired by the recent experiment by Kazimierczuk \emph{et
al}.~\cite{Kazimierczuk}. We have calculated the absorption
spectrum, obtaining a very good agreement between the calculated
and the experimentally observed spectra, including the splitting
between the $P$ and $F$ excitons. We also obtained a good
agreement with the experimental and theoretical results by Thewes
et al.~\cite{Thewes}. We have derived the dipersion relation for
polariton waves, which differs from the analogous relations for
other semiconductors as, for example, GaAs. All these interesting
features of  excitons with high $n$ number which are examined  and
discussed  on the basis on our theory might possibly provide deep
insight into the nature of Rydberg excitons in solids and provoke
their application to design all-optical flexible switchers and
future implementation in quantum information processing.
\section*{Acknowledgement}
The authors ave very indebted to M. A. Semina and M. M. Glazov for
their remarks and valuable discussion.

\appendix
\section{Appendix A. Derivation of the transition dipole density}
\label{Appendix A}

In the real density matrix approach the coupling between the band
edge and the radiation field is described by a smeared out
transition dipole density $\textbf{M}(\textbf{r})$, $\textbf{r}$
being the relative electron-hole distance. For the transition
between the subbands $\nu_1, \nu_2$ the definition is \cite{StB87}
with the momentum operator $\textbf{p}_\rho(\textbf{k})$

\begin{eqnarray}\label{mdipol1} {\bf M}^{\nu_1\nu_2}({\bf
r})&=&:{\bf M}_{\rho}({\bf r})=
\int\limits_{1.{\rm BZ}}{\rm d}^3 k\tilde{M}_{\rho}({\bf k})e^{{\rm i}{\bf k}{\bf r}}\nonumber\\
&=&\frac{1}{(2\pi)^3}\frac{{\rm i}e\hbar}{m_0} \int\limits_{1.{\rm
BZ}}{\rm d}^3 k\frac{{\bf p}_{\rho}e^{{\rm i}{\bf k}{\bf
r}}}{E_c({\bf k})-E_{v\rho}({\bf k})}.
\end{eqnarray}
For allowed interband transitions we assume that
$\textbf{p}(\textbf{k})\approx \textbf{p}(0)$ and that the
interband energy is parabolic in relevant parts of
$\textbf{k}$-space
\begin{equation}
E_c(\textbf{k})-E_v(\textbf{k})=E_g+\frac{\hbar^2k^2}{2\mu}.
\end{equation}
Extending the integration to entire $\textbf{k}$-space one obtains
(\cite{StB87}, \cite{CzB85}, see also \cite{Linear})
\begin{equation}\label{zerothorder}
\textbf{M}(r)=\textbf{M}_0\frac{1}{4\pi r_0^2r}e^{-r/r_0},
\end{equation}
with $r_0$ defined in eq. (\ref{r0}). In the case of the forbidden
transtions $p(0)$ vanishes. The relevant transtion dipole density
will be obtained from (\ref{mdipol1}) by expansing the momentum
operator in powers of $\textbf{k}$. To lowest nonvanishing order
in $\textbf{k}$, we have $\textbf{p}(\textbf{k})\propto
\textbf{k}$. Note that $k$ will arrive in the numerator of
(\ref{mdipol1}) after applying on both sides the operator
$\hbox{\boldmath$\nabla$}$. So we obtain
\begin{equation}
\textbf{M}(\textbf{r})\propto
\,\hbox{\boldmath$\nabla$}\frac{e^{-r/r_0}}{r}=\frac{\textbf{r}}{r^3r_0}(r+r_0)e^{-r/r_0}.
\end{equation}
Requiring the normalization of the radial part we obtain the
formula (\ref{gestoscwzbronione}).

In general, the momentum operator $\textbf{p}_{cv}(\textbf{k})$
can be expanded in series of $k$:
\begin{equation}
\textbf{p}_{cv}(\textbf{k})=\textbf{p}_{cv}(0)+\textbf{k}\left[\hbox{\boldmath$\nabla$}_k
\textbf{p}_{cv}(\textbf{k})\right]_{k=0}+....
\end{equation}
Since we exploited the terms with $\textbf{p}_{cv}(0)$ and the
term $\propto k$, the next terms will be proportional to
$\textbf{p}(\textbf{k})\propto \textbf{k}(\textbf{k}\cdot
\textbf{k})$, $\textbf{k}\cdot \textbf{k}$ means dyadic product.
Performing, as above, the derivative operation with respect to
$z$, (or $x,y$, respectively) on (\ref{zerothorder}) and retaing
the largest contributions, we obtain (for the sake of
exemplification, we present the $z$-component)
\begin{eqnarray}
M_2&\propto&
\frac{3z^2-r^2}{r^5}e^{-r/r_0}=\sqrt{\frac{16\pi}{5}}\frac{Y_{20}}{r^2}e^{-r/r_0}
=\sqrt{\frac{16\pi}{5}}r_0\frac{Y_{20}}{r^2}e^{-r/r_0},\\
\label{Fexcitons}M_{3z}\propto \frac{\partial^3}{\partial z^3}&&=\frac{3z(3r^2-5z^2)}{r^7}e^{-r/r_0}=-12\sqrt{\frac{\pi}{7}}\frac{Y_{30}}{r^4}e^{-r/r_0},\\
M_{3x}\propto \frac{\partial^3}{\partial
x^3}&&=\frac{3x(3r^2-5x^2)}{r^7}e^{-r/r_0}\nonumber\\&&=\frac{1}{r^4}\left[\sqrt{\frac{3\pi}{7}}\left(Y_{31}-Y_{3-1}\right)-
\sqrt{\frac{5\pi}{7}}\left(Y_{33}-Y_{3-3}\right)\right]e^{-r/r_0},\nonumber\\
M_{3y}\propto \frac{\partial^3}{\partial
y^3}&&=\frac{3y(3r^2-5y^2)}{r^7}e^{-r/r_0}\nonumber\\&&=-{\rm
i}\frac{1}{r^4}\left[\sqrt{\frac{3\pi}{7}}\left(Y_{31}+Y_{3-1}\right)-
\sqrt{\frac{5\pi}{7}}\left(Y_{33}+Y_{3-3}\right)\right]e^{-r/r_0},
\end{eqnarray}
where $M_2$ corresponds to $D$ excitons, and $M_3$ to $F$
excitons, $M_3$ in unnormalized form. \setcounter{equation}{0}

\section{The calculation of the coefficients $b_{n1}$ and $b_{n3}$}
\label{Appendix B} First we calculate the coefficients $b_{n1}$
which define the susceptibility related to the $P$-excitons. Using
the relations (\ref{coefficients1}), (\ref{wodorradial}), and
(\ref{normalizacja_atom_wodoru}), we obtain
\begin{eqnarray}\label{normalizacja_atom_wodoru2}
C_{n1}^2\cdot\left(\frac{2}{na^*}\right)^2&=&
\left(\frac{1}{3!}\right)^2\frac{(n+1)!}{2n(n-2)!}\left(\frac{2}{na^*}\right)^{3}\cdot\left(\frac{2}{na^*}\right)^2\nonumber\\
&=&\frac{4}{9}\frac{n^2-1}{n^5}\frac{1}{a^{*5}}.
\end{eqnarray}
The above expression will be used in the calculation of the
coefficients $b_{n1}$
\begin{eqnarray}
&&b_{n1}=\frac{8\pi}{3}\left(\int\limits_0^\infty {\rm d}r r^2
R_{n1}(r)M(r)\right)^2\nonumber\\
&&=\frac{4\pi}{3}M_{10}^2C_{n1}^2\cdot\left(\frac{2}{na^*}\right)^2\left[\int\limits_0^\infty{\rm
d}r\,r\,M\left(-n+2,4,\frac{2r}{a^*n}\right)\exp\left(-\frac{r}{na^*}\right)\frac{r+r_0}{r_0^2}e^{-r/r_0}\right]^2\\
&&=\frac{4\pi}{3r_0^2}M_{10}^2C_{n1}^2\cdot\left(\frac{2}{na^*}\right)^2\left\{\int\limits_0^\infty{\rm
d}r\,r(r+r_0)\,M\left(-n+2,4,\frac{2r}{a^*n}\right)\exp\left[-r\left(\frac{1}{na^*}+\frac{1}{r_0}\right)\right]\right\}^2.\nonumber
\end{eqnarray}
Let us first calculate the integrals involving the confluent
hypergeometric function
\begin{eqnarray*}
&&I=\int\limits_0^\infty{\rm
d}r\,r(r+r_0)\,M\left(-n+2,4,\frac{2r}{a^*n}\right)\exp\left[-r\left(\frac{1}{na^*}+\frac{1}{r_0}\right)\right]\nonumber\\
&&=\int\limits_0^\infty{\rm
d}r\,r^2\,M\left(-n+2,4,\frac{2r}{a^*n}\right)\exp\left[-r\left(\frac{1}{na^*}+\frac{1}{r_0}\right)\right]\\
&&+r_0\int\limits_0^\infty{\rm
d}r\,r\,M\left(-n+2,4,\frac{2r}{a^*n}\right)\exp\left[-r\left(\frac{1}{na^*}+\frac{1}{r_0}\right)\right]\nonumber.
\end{eqnarray*}
The above integrals can be expressed in terms of the
hypergeometric function:
\begin{equation}
J_{\alpha\gamma}^\nu=\int\limits_0^\infty \;e^{-\lambda z}z^\nu
M(\alpha,\gamma,kz)\,{\rm
d}z=\Gamma(\nu+1)\lambda^{-\nu-1}F\left(\alpha,\nu+1,\gamma,\frac{k}{\lambda}\right),
\end{equation}
$F(\alpha,\beta,\gamma,z)$ being the hypergeometric series
\begin{equation}
F(\alpha,\beta,\gamma,z)=1+\frac{\alpha\beta}{\gamma}\,\frac{z}{1!}+\frac{\alpha(\alpha+1)\beta(\beta+1)}{\gamma(\gamma+1)}\,\frac{z^2}{2!}+\ldots.
\end{equation}
In the first approximation, assuming that $r_0\ll na^*$, which is
certainly correct for large values of $n$, we have
\begin{eqnarray*}I &&\approx
2\left(\frac{na^*r_0}{r_0+na^*}\right)^3+r_0\left(\frac{na^*r_0}{r_0+na^*}\right)^2=\left(\frac{na^*r_0}{r_0+na^*}\right)^2
\left[2\frac{na^*r_0}{r_0+na^*}+r_0\right]\nonumber\\
&&=r_0(3na^*+r_0)\frac{(na^*r_0)^2}{(r_0+na^*)^3}
\end{eqnarray*}
Thus
\begin{eqnarray}\label{coefficientsbn}
&&b_{n1}=\frac{4}{9}\frac{n^2-1}{n^5}\frac{1}{a^{*5}}\frac{4\pi}{3r_0^4}M_{10}^2\left[3\left(\frac{na^*r_0}{r_0+na^*}\right)^3\right]^2\nonumber\\
&&=M_{10}^2\frac{n^2-1}{n^5}\frac{16\pi}{3}\left(\frac{a^*}{r_0}\right)^4\frac{1}{a^{*3}}\left(\frac{nr_0}{r_0+na^*}\right)^6.
\end{eqnarray}
For the $F$ excitons, taking $\ell=3$, we have
\begin{eqnarray}
&&b_{n3}=\frac{144\pi}{7}\left(\int\limits_0^\infty {\rm d}r r^2
R_{n3}(r)M_3(r)\right)^2\nonumber\\
&&=\frac{144\pi}{7}M_{30}^2C_{n3}^2\cdot\left(\frac{2}{na^*}\right)^6\left\{\int\limits_0^\infty{\rm
d}r\,r\,M\left(-n+4,8,\frac{2r}{a^*n}\right)\exp\left[-r\left(\frac{1}{na^*}+\frac{1}{r_0}\right)\right]\right\}^2\nonumber\\
&&=\frac{256\cdot
144\cdot\pi\,r_0^2}{7(7!)^2a^{*9}}M_{30}^2\frac{(n^2-9)(n^2-4)(n^2-1)}{n^9}.
\end{eqnarray}
Roughly speaking, the oscillator strengths related to $F$ excitons
also scale as $n^{-3}$, but are of order in magnitudes smaller
than those of $P$ excitons, due to the factor $(7!)^2$ in
denominator. In addition, the factor $(n^2-9)(n^2-4)(n^2-1)/{n^9}$
is smaller than $1/n^3$ as we can see, for example, for $n=4$
(0.0156 compared to $4.8\cdot 10^{-3}$). The same holds for the
next odd $\ell$ value (5), where the factor $(11!)^2$ arrives in
the denominator of the oscillator strength.

\section{Estimation of the coherence radius and the transition dipole matrix elements
$M_0$}\label{Appendix C}

The formula (\ref{polariton}) allows to estimate the matrix
element $M_{10}$. Since $k=0$ at a longitudinal frequency
$\omega_L$, taking from the r.h.s. the main contribution at $n=2$,
we have
\begin{eqnarray}
\epsilon_0\epsilon_b\Delta_{LT}^{(2)}&=&2b_{21},\\
\Delta_{LT}^{(2)}&=&\hbar\omega_{L2}-\hbar\omega_{T2},\\
\label{delta2}\Delta_{LT}^{(2)}&=&\frac{\pi}{\epsilon_0\epsilon_b
a^{*3}}M_{10}^2\left(\frac{a^*}{r_0}\right)^4\left(\frac{2r_0}{r_0+2a^*}\right)^6,\nonumber\\
&=&R^*\cdot
2\frac{2\mu}{\hbar^2}\frac{M_{01}^2}{\pi\epsilon_0\epsilon_ba^*}f(r_0,a^*),\end{eqnarray}
with
\begin{equation}
f(r_0,a^*)=\frac{\pi^2}{2}\left(\frac{a^*}{r_0}\right)^4\left(\frac{2r_0}{r_0+2a^*}\right)^6,
\end{equation}
and with regard to the relation $2\mu/\hbar^2=1/R^*a^{*2}$. Higher
order coefficients $b_{n1}$ can be expressed in terms of $b_{21}$
\begin{equation}\label{higherbn}
{b_{n1}}=\frac{32(n^2-1)}{3n^5}\left[\frac{nr_0(r_0+2a^*)}{2r_0(r_0+na^*)}\right]^6{b_2}=\frac{16(n^2-1)}{3n^5}\left[\frac{nr_0(r_0+2a^*)}
{2r_0(r_0+na^*)}\right]^6\epsilon_0\epsilon_b\Delta_{LT}^{(2)}=\frac{1}{2}f_{n1}\epsilon_0\epsilon_b\Delta^{(2)}_{LT},
\end{equation}
with
\begin{equation}\label{oscillatorforces}
f_{n1}=\frac{32(n^2-1)}{3n^5}\left[\frac{nr_0(r_0+2a^*)}
{2r_0(r_0+na^*)}\right]^6.
\end{equation}
The higher order longitudinal-transversal energies
$\Delta_{LT}^{(n)}=\hbar\omega_{Ln}-\hbar\omega_{Tn}$ are related
to the coefficients $b_{n1}$ by
\begin{equation}
2b_{n1}=\epsilon_0\epsilon_b\Delta_{LT}^{(n)}.
\end{equation}
When $r_0\ll a^*$, (\ref{higherbn}) becomes
\begin{equation}
b_{n1}=\frac{32}{3}\,\frac{n^2-1}{n^5}\,b_{21},
\end{equation}
or, in terms of $\Delta_{LT}^{(2)}$,
\begin{equation}
b_{n1}=\frac{16}{3}\,\epsilon_0\epsilon_b\frac{n^2-1}{n^5}
\Delta_{LT}^{(2)},\qquad n=3,4,\ldots.
\end{equation}
In the above equations and definitions, two parameters are used:
the transition dipole matrix element $M_{01}$ and the coherence
radius $r_0$.  As it follows from eq. (\ref{delta2})
\begin{eqnarray}
\frac{\Delta_{LT}^{(2)}}{R^*}&=&
2\frac{2\mu}{\hbar^2}\frac{M_{01}^2}{\pi\epsilon_0\epsilon_ba^*}f(r_0,a^*)=2\frac{M_{01}^2}{\epsilon_0\epsilon_b\pi
R^*a^{*3}}f(r_0,a^*),\end{eqnarray} the longitudinal-transversal
splitting energy and the coherence radius are not independent
quantities. Take, for example, the value of the transition dipole
matrix element given in Ref. \cite{Mysyrowicz} for the $1s\to 2p$
transition, in our notation $M_0=4.6~e\AA=1.08\cdot
10^{-28}~\hbox{C m}$. Using this value and other parameters for
Cu$_2$O we obtain
\begin{equation}
\frac{\Delta_{LT}^{(2)}}{R^*}=4.55 f(r_0,a^*),
\end{equation}
which means that a value of $\Delta_{LT}^{(2)}$ the order of 1 meV
will be obtained for a small (compared to $a^*$) value of $r_0$.
Thus the situation is the following: either we know the exact
value of the LT-splitting energy (as is the case, for example, for
GaAs), and the coherence radius can be estimated, or we consider
$\Delta_{LT}$ and $r_0$ as two unknown quantities. Then we need
two equations for establish them. Such equations can be obtained,
for example, from the behavior of the dielectric function (here in
scalar notation) $\epsilon(\omega)$,
\begin{eqnarray}
\frac{c^2k^2}{\omega^2}&=&\epsilon(\omega,k)=\epsilon_b+\chi(\omega,k),
\end{eqnarray}
from which we obtain two equations
\begin{eqnarray}\label{estimationrzero}
0&=&\epsilon_b+\chi(\omega_L,0),\\
\epsilon_\infty&=&\epsilon_b+\chi(\omega_g,0),
\end{eqnarray}
($\epsilon_\infty = 6.5$ for Cu$_2$O). From the above relations,
which, in turn, also depend on the number of states taken into
account, the relation between $r_0$ and $\Delta_{LT}$ can be
obtained.

 \end{document}